\documentstyle[preprint,aps,fleqn]{revtex}

\begin{document}

\draft

\title{Constraints on QCD Sum-rules from  the H\"older Inequalities}
\author{M. Benmerrouche$^a$, G. Orlandini$^b$, T.G. Steele$^a$ }
\address{$^a$Department of Physics \& Engineering Physics and \\
Linear Accelerator Laboratory\\
University of Saskatchewan\\
Saskatoon, Saskatchewan S7N 0W0, Canada\\[1em]
$^b$Dipartimeto di Fisica, Universit\`a degli Studi di Trento\\
I-38050 Povo (Trento), Italy \\
and Istituto Nazionale di Fisica Nucleare, Gruppo collegato di Trento}

\date{\today}
\maketitle

\begin{abstract}
A new technique based on H\"older's integral inequality is applied to QCD
sum-rules to provide fundamental constraints on the sum-rule parameters.
These constraints must be satisfied if the sum-rules are to consistently
describe integrated physical cross-sections, but
these constraints do not require any experimental data and therefore can be
applied to any hadronic spectral function. As an illustration of this
technique the Laplace
sum-rules of the light-quark correlation function for the
vector and the axial-vector currents are examined in detail.
We find examples of inconsistency
between the inequalities and sum-rule parameters
used in some previous analyses of the vector and axial-vector channels.
\end{abstract}

\vspace{0.2in}

QCD sum-rules\cite{svz,svz2,rry,nar} have demonstrated their utility in
numerous theoretical determinations of hadronic properties.  In this
approach the QCD condensates  parametrize nonperturbative aspects
of the vacuum, and  through the operator-product expansion\cite{wils},
the condensates generate power-law corrections to correlation
functions of hadronic currents. These power-law contributions are
absent in a purely perturbative calculation and are an essential
feature of the sum-rules used to determine hadronic properties.

Despite the success of QCD sum-rules, there remain several fundamental issues
concerning their application.  In particular, the values of the QCD
condensates,
applicability and implementation of the continuum (duality) hypothesis,
and the energy range in which the sum-rules are reliable are significant
issues in the use of QCD sum-rules in hadronic physics.
One of these issues is well illustrated by the dimension-six quark
condensate in (light-quark) vector current correlation functions
where estimates differ by factors of $2$ or
more\cite{svz,eidel,laun,bert,dom,gim1,gim2}.

In this paper we
will present a method based on H\"older's integral inequality which
provides fundamental constraints on the QCD sum-rules.  These constraints must
be satisfied if the sum-rules are to consistently describe integrated physical
cross-sections. Using this technique, non-trivial information relating the
continuum threshold, sum-rule energy scale, and QCD condensate parameter space
will be obtained.
These constraints then provide insight into the issues  concerning the
continuum hypothesis and the  energy range in which the sum-rules are reliable.
Although Schwarz and H\"older inequalities have
been studied in connection with lattice gauge theories to
demonstrate some general properties of the hadronic
spectrum\cite{latt1,latt2,latt3} they have not previously been applied
to the QCD sum-rules.

The H\"older inequality technique will be illustrated by the
Laplace sum-rules
 involving the light quark vector current
(related to the $\rho$) and light quark axial vector current (related to the
$A_1$).  These channels have been  chosen because they have been
extensively studied, particularly for the vector channel where the analysis
of the $\rho$ meson has become a paradigm for sum-rule techniques.

H\"older's inequality\cite{hold,hold2} for integrals defined over a measure
$d\mu$ is

\begin{eqnarray}
\biggl|\int_{t_1}^{t_2} f(t)g(t) d\mu \biggr| &\le&
\left(\int_{t_1}^{t_2} \big|f(t)\big|^ p d\mu \right)^{1/p}
\left(\int_{t_1}^{t_2} \big|g(t)\big|^q d\mu \right)^{1/q}, \nonumber \\
&& \label{holder} \\[-1em]
\frac{1}{p}+\frac{1}{q} &=&1~;\quad p,~q\ge 1\quad . \nonumber
\end{eqnarray}
When $p=q=2$ the H\"older inequality reduces to the well known Schwarz
inequality. The key idea in applying H\"older's inequality to sum-rules is
recognizing that for a typical correlation function $\Pi(Q^2)$,
$Im\, \Pi(q^2)$ is positive because of its relation to physical
cross-sections and can thus serve as the measure
$d\mu=Im\,\Pi(t) dt$ in (\ref{holder}).

Laplace sum-rules are also related to $Im\, \Pi(t)$ through a Borel
transform of a dispersion relation
\begin{equation}
{\cal R}_k(\tau,s_0)=\int_{t_0}^{s_0}\,Im\, \Pi(t) t^k e^{-t\tau}dt
\qquad ,\quad k={\rm integer},
\label{sr}
\end{equation}
where $t_0$ is a physical threshold and $s_0$ is the continuum
representing the minimum energy needed for local duality\cite{dual}.
In the sum-rule method the QCD contributions to ${\cal R}_k(\tau, s_0)$
on the left hand side of (\ref{sr}) are used to extract the
phenomenological content of $Im\Pi(t)$.
Among other issues, the applicability of the QCD continuum hypothesis (duality)
used to model $Im\,\Pi(t)$ above the energy scale $s_0$ can be examined through
the H\"older inequalities.

Returning to (\ref{holder}) with $d\mu=Im\Pi(t) dt$, $f(t)=t^\alpha
e^{-at\tau}$, $g(t)=t^\beta e^{-bt\tau}$ and appropriate
integration limits we find
\begin{equation}
{\cal R}_{\alpha+\beta}(\tau,s_0)\le
{\cal R}^{1/p}_{\alpha p}(ap\tau, s_0)
{\cal R}^{1/q}_{\beta q}(bq\tau, s_0)
\;;\quad a+b=1\quad .
\label{rat1}
\end{equation}
Imposing restrictions that we have the integer values $k$ needed for
the sum-rules (\ref{sr}) leads to the following set of inequalities.
\begin{eqnarray}
{\cal R}_0[\omega\tau_{min}+(1-w)\tau_{max},s_0] &\le&
{\cal R}^\omega_0[\tau_{min},s_0] {\cal R}^{1-\omega}_0[\tau_{max},s_0],
\label{ineqa} \\
{\cal R}_1[\omega\tau_{min}+(1-w)\tau_{max},s_0] &\le&
{\cal R}^\omega_1[\tau_{min},s_0] {\cal R}^{1-\omega}_1[\tau_{max},s_0],
\label{ineqb}\\
{\cal R}_1[\frac{\tau_{min}+\tau_{max}}{2},s_0] &\le&
{\cal R}_2^{1/2}[\tau_{min},s_0]{\cal R}_0^{1/2}[\tau_{max},s_0],
\label{ineqc}\\
{\cal R}_1[\frac{\tau_{min}+\tau_{max}}{2},s_0] &\le&
{\cal R}_2^{1/2}[\tau_{max},s_0]{\cal R}_0^{1/2}[\tau_{min},s_0],
\label{ineqd}\\
0\le\omega\le 1~&;&\quad\tau_{min}\le\tau_{max}
\end{eqnarray}
Similar inequalities can be obtained for higher sum-rules with
$k\ge 2$.
However,
as $k$ increases,
the leading QCD condensate contributions to the
sum-rules begins to depend upon poorly-understood high dimension condensates,
 so our analysis will concentrate upon (\ref{ineqa}) and
(\ref{ineqb}) where $k<2$.
Furthermore, for small $\tau_{max}-\tau_{min}$ (\ref{ineqc}) and (\ref{ineqd})
are in principle contained in the first two inequalities.
Thus the following ratios reflecting the inequalities (\ref{ineqa}) and
(\ref{ineqb}) will be used to study the  parameter space of the QCD sum-rules.
\begin{eqnarray}
\rho_0 &\equiv&\frac{{\cal R}_0[\omega\tau_{min}+(1-
\omega)\tau_{max},s_0]}{{\cal R}_0^\omega[\tau_{min},s_0]
{\cal R}_0^{1-\omega}[\tau_{max},s_0]} \le 1
\label{rat2a}\\
\rho_1 &\equiv&\frac{{\cal R}_1[\omega\tau_{min}+(1-
\omega)\tau_{max},s_0]}{{\cal R}_1^\omega[\tau_{min},s_0]
{\cal R}_1^{1-\omega}[\tau_{max},s_0]}\le 1
\label{rat2b}
\end{eqnarray}

In summary, if the QCD sum-rules are a valid and consistent representation of
the integration of $Im\Pi(t)$ in (\ref{sr}) then  the sum-rules
${\cal R}_k(\tau,s_0)$ must satisfy the above H\"older inequalities.

To analyze the implications of these inequalities, the QCD
predictions for the sum-rules are needed.  For the light-quark
vector and axial\footnote{For the axial current this
represents the transverse projection of the correlation function.}
currents the results (to two-loops in perturbative corrections, leading order
in QCD condensates) are \cite{svz,rry}
\begin{eqnarray}
8\pi^2{\cal R}_0[\tau,s_0]&=&\frac{1}{\tau}
(1+\frac{\alpha(1/\tau)}{\pi})[1-e^{-s_0\tau}] +C_2
+C_4\langle O_4 \rangle\tau \nonumber \\
&+&\frac{1}{2}C_6\langle O_6\rangle\tau^2 + \frac{1}{3!}C_8\langle
O_8\rangle \tau^3 + {\rm higher~dimension~condensates}
\label{vasra}\\
8\pi^2{\cal R}_1[\tau,s_0]&=&\frac{1}{\tau^2}
(1+\frac{\alpha(1/\tau)}{\pi})[1-(1+s_0\tau)e^{-s_0\tau}]
-C_4\langle O_4\rangle \nonumber \\
&-&C_6\langle O_6\rangle \tau -\frac{1}{2}C_8\langle O_8\rangle \tau^2
+{\rm higher~dimension~condensates}
\label{vasrb}\\
C_4\langle O_4\rangle&=&\frac{\pi}{3}\langle \alpha G^2\rangle+8\pi^2
m\langle \bar q q\rangle\quad{\rm vector}
\label{vasrc}\\
C_4\langle O_4\rangle &=&\frac{\pi}{3}\langle \alpha G^2\rangle -8\pi^2 m
\langle \bar q q\rangle\quad{\rm axial~ vector}
\label{vasrd}\\
C_6\langle O_6\rangle &=&-4\pi^3\frac{224}{81}\alpha
(\langle \bar q q\rangle)^2 \quad{\rm vector}
\label{vasre}\\
C_6\langle O_6\rangle &=&44\pi^3\frac{32}{81}\alpha (\langle \bar q q\rangle)^2
\quad{\rm axial~ vector}
\label{vasrf}
\end{eqnarray}
where the vacuum saturation hypothesis\cite{svz}
has been used for the
dimension-six condensates and the (small) perturbative contribution
from quark masses has been neglected.
For brevity, we have not explicitly shown the dimension-eight
operators and refer instead to the literature\cite{gim1,dim8,dim82}.
Finally, although there are no vacuum condensates of dimension two,
the phenomenological possibility of such  contributions
(apart from the small quark mass corrections)
has been suggested in the context of renormalons \cite{domc2}.
Such non-OPE corrections are represented by the constant term $C_2$, and
their effect will be investigated as part of the sum-rule parameter space.

The gluon condensate is now reasonably well established\cite{gim1}
to lie within the range
$\langle \alpha G^2 \rangle = 3\;(0.050 \pm 0.015)/\pi\,{\rm GeV}^4$.
However, it has been suggested that the vacuum saturation hypothesis
\cite{svz}, leading to
$C_6\langle O_6\rangle =-0.06\,{\rm GeV}^6$ (vector) and
$C_6\langle O_6\rangle =\frac{11}{7} 0.06\,{\rm GeV}^6$ (axial),
underestimates the magnitude
of the dimension-six quark condensate by a factor of $2$ or more
\cite{eidel,laun,bert,dom,gim1,gim2}.
The many dimension-eight operators fall into two distinct classes
consisting of operators amenable to estimation through the vacuum saturation
hypothesis and fermionic operators consisting of contractions of
$\langle \bar q D_\mu D_\nu D_\lambda D_\rho D_\omega q\rangle$.  These
fermionic condensates have been estimated at lower dimension
\cite{dim8} by assuming that quarks
have a virtuality of $M^2\approx 0.3 \,{\rm GeV}^2$ which replaces each
covariant derivative with a mass scale $M$.  Using these ideas we estimate
that the non-fermionic
condensates dominate $C_8\langle O_8\rangle$ leading to a result of
$C_8\langle O_8\rangle \approx 4\times 10^{-3} \,{\rm GeV}^8$.
However, the analysis of \cite{dom,gim1,gim2} suggests large deviations
from this
value. This does not necessarily reflect a complete failure of the vacuum
saturation hypothesis since even a 10\% deviation from vacuum saturation for
each individual operator can
accumulate through the combination of the many dimension-eight condensates.

To analyze the inequalities (\ref{rat2a},\ref{rat2b}), we restrict the
parameter space by performing a local analysis with
$\tau_{max}-\tau_{min}=\delta\tau=0.01\,{\rm GeV}^{-2}$ and setting
$\Lambda_{MS}=0.15\,{\rm GeV}$.  Further decrease and moderate increase in
the value of $\delta\tau$ does not affect the conclusions presented below,
and the effects of changing $\Lambda_{MS}$ are negligible.
The QCD condensates are then fixed to a particular set of values and the
regions
of $s_0$, $\tau$ parameter space leading to $\rho_0<1$ and $\rho_1<1$ for
all $0<\omega<1$ are determined.  The values of the condensates are then
varied and the process is repeated.

The results of this analysis for both the vector and axial-vector channels are
illustrated in the Figures, corresponding
to specific values of the condensates used or determined in the literature.
In each figure the shaded region
represents the admissible $s_0$, $\tau$ parameter space where the
inequalities are satisfied.  The boxed regions in the figures are the
values of $s_0$ and $\tau$ range given in the literature. As is
evident from the figures there are several cases where the parameters
used in a particular sum-rule analysis are inconsistent with the inequalities.
To determine whether this inconsistency is significant we have analyzed the
sum-rules to determine how the uncertainties inherent to the sum-rule
technique (such as truncation of the OPE beyond condensates of a certain
dimension) affect the inequalities\footnote{We are grateful to the referee
for this suggestion.}.
This has been done for Figures 1-3 by assuming that the power-law corrections
in ${\cal R}_0$ and
${\cal R}_1$ have an intrinsic error of 10\% at $1.0\,{\rm GeV}$.
This is a generous
estimate of the uncertainty by comparison with the assumptions of the
standard sum-rule error analysis \cite{svz,svz2,rry}
which leads to less than a 1\% error in the power-law corrections at
$1.0\,{\rm GeV}$.
Our uncertainty is then modelled by a condensate representing the first
truncated term in the OPE,
and then considering values of the condensate which have a 10\% effect in the
power law
corrections at $1.0\,{\rm GeV}$.  Clearly other error models could be chosen
but for the purpose of this
work we wish to emphasize the method based on the inequalities.
For Figure 4, an
alternative approach for studying the effects of truncation in the OPE
will be discussed below.

The details of the individual figures vary, but two common features
persist:  the existence of a lower bound on the continuum threshold
$s_0$ and an upper bound on the sum-rule energy parameter $\tau$.
An interesting feature of the bound on $s_0$ is that it is generally
smaller in
the vector channel than in the axial channel, in agreement with the
trend observed in phenomenology.

In Fig. \ref{figure1} the allowed $s_0$-$\tau$ parameter space for the vector
channel is shown for three sets of the condensates.
In all three cases the dimension-eight and higher condensates are ignored.
The bottom graph corresponds to the standard Shifman-Vainshtein-Zakharov
values of the condensates \cite{svz,svz2}, the middle graph incorporates a
larger value of the gluon condensate, and the top graph represents twice
vacuum saturation for the dimension six-condensate \cite{nar}.
The square boxes represent the range of values for
$s_0$
and $\tau$ used in the literature for the sum-rule analysis corresponding
to the condensates
\footnote{In analyses where no range was reported for $s_0$ a range of
$0.5 \, {\rm GeV}^2$ has been assumed.} .
The dashed line represents the border of the parameter
space when a dimension-two phenomenological contribution of
$C_2=-0.09\,{\rm GeV}^2$ is included as suggested by the upper bounds in
\cite{domc2}.
The dotted lines represent the border of the parameter space consistent with
the inequalities after including the effect of uncertainty in the power-law
corrections.
For the lower two graphs (standard values) the shaded area
(and its extension to the dotted line after including uncertainties)
overlaps
with a large portion of the boxed region, so the sum-rule analysis\cite{svz2}
is consistent with the inequalities.  By contrast, the upper graph has a very
small overlap between
the shaded region (with its extension to the dotted line) and boxed region so
there is a minimal consistency between the inequalities and sum-rule
parameters.
In general,
Fig. \ref{figure1} shows that the vector channel parameter space consistent
with the inequalities  lies within the bounds $s_0>1.0\,{\rm GeV}^2$,
$\tau<1.8\,{\rm GeV}^{-2}$
for the standard values  and $s_0>1.5\,{\rm GeV}^2$,
$\tau<1.4\,{\rm GeV}^{-2}$ for twice vacuum saturation.

Fig. \ref{figure2} represents a similar analysis for three sets of
condensates in
the axial-vector channel.  In all cases the dimension-eight and higher
condensates are ignored.  The bottom two graphs correspond to standard values
of the condensates as
used in \cite{rry} with the middle graph using a slightly larger gluon
condensate.
The top graph again corresponds to twice vacuum saturation for the
dimension-six condensate\cite{nar}. All other features are the same as in
Fig. \ref{figure1}.  We see from Fig. \ref{figure2}
that in all cases the sum-rule analyses of the axial vector channel are
inconsistent with the inequalities even when the effect of uncertainties are
considered.
Fig. \ref{figure2} shows that the axial-vector channel parameter space
consistent with the inequalities  lies within the bounds
$s_0>2.5\,{\rm GeV}^2$, $\tau<1.1\,{\rm GeV}^{-2}$
for the standard values  and $s_0>3.0\,{\rm GeV}^2$,
$\tau<0.8\,{\rm GeV}^{-2}$ for twice vacuum saturation.

It is evident from both Fig. \ref{figure1} and Fig. \ref{figure2}
that the inequalities are insensitive to a reasonable variation
in the gluon condensate, but rather sensitive to the value of the
dimension-six condensate.
The dimension-two phenomenological condensate also has a relatively minor
effect on the parameter space consistent with the inequalities.

In Fig. \ref{figure3} and \ref{figure4} we repeat the procedure including the
effects of higher dimension condensates. The value of dimension eight
condensates were determined for both the vector and axial-vector channels
by Dominguez and S\`{o}la\cite{dom} using
finite energy (FESR) and Laplace sum-rules. The dimension-eight condensates
were found to improve duality between
the experimental data and QCD and their values hint to a possible larger
fermionic contribution to $C_8\langle O_8 \rangle$. In Fig. \ref{figure3}, we
display our results based on the average values for the
higher dimension
condensates as given
in Table 1 of \cite{dom}. As in the other figures, we also show (square boxes)
the ranges of
values for $s_0$ and $\tau$ employed in this same work \cite{dom}.
For the vector
channel, the shaded area (and its extension to the dotted line after
including uncertainties)
does not overlap with the box and therefore the values of the condensates are
{\em inconsistent} with the $s_0-\tau$ region as analysed in
Dominguez-S\`{o}la work. The admissible parameter space lies within the
bounds $s_0 \ge 2.3 \,{\rm GeV}^2$ and
$\tau \le 0.9\, {\rm GeV}^{-2}$ for the vector channel while
$s_0 \ge 0.5 \,{\rm GeV}^2$
and $\tau \le 1.3 \, {\rm GeV}^{-2}$ for the axial channel.

In Fig. \ref{figure4}, we display the results
based on the value of the condensates as given in Table 4 of \cite{gim2}
which contains condensates up to dimension sixteen.
The shaded region is the $s_0-\tau$ parameter space consistent with
the inequalities using the condensates up to dimension sixteen.
Since these values of the condensates are reasonably consistent with \cite{dom}
for dimension 8 and less, the effect of truncating the OPE can be
explicitly studied
in this case by omitting the condensates above dimension 8 resulting in a shift
of the border of the parameter space to the dotted line.
The effect in this case is more significant than the error model considered
in the other figures
because the contribution of condensates of dimension 10 to 16 is significantly
more than  10\% at $1.0 \,{\rm GeV}$.
It is interesting that
including higher dimension condensates does not necessarily increase the region
consistent with the inequlaities  as evidenced by the axial vector channel.
 As in the other figures, the boxed regions show the
$s_0-\tau$ interval used in this same work \cite{gim2}.
Clearly there are regions of $s_0-\tau$ parameter space used in the analysis
\cite{gim2} which are inconsistent with the inequalities.
The admissible parameter space lies within the bounds $s_0 \ge 1.6 \,
{\rm GeV}^2$ and
$\tau \le 0.95\, {\rm GeV}^{-2}$ for the vector channel while $s_0 \ge 2.2 \,
{\rm GeV}^2$ and $\tau \le 1.1 \, {\rm GeV}^{-2}$ for the axial channel.

The inequalities should be viewed as a test of both the
validity of the
continuum hypothesis and of the upper bound on $\tau$ (lowest energy) beyond
which the neglected or unknown effects in the sum-rule become substantial.
In general, features of the allowed parameter space that are independent of
$s_0$  (such as the rising vertical sections) represent the upper range on
$\tau$ for which the sum-rule becomes unreliable, and the lower horizontal
portions suggest a failure of the continuum hypothesis regardless of the
energy scale $\tau$.
It is significant that the bounds on the $s_0$ and $\tau$ parameter space
are  obtained only by demanding that the
sum-rule be consistent with its phenomenological description in terms of
an integrated cross-section through $Im \,\Pi(t)$, leading to the
H\"older inequality constraint.  This should be contrasted with the
conventional approach of determining an upper bound on $\tau$
where an assumption on the uncertainties in the power-law corrections is
made, and the limit on $\tau$ represents an energy at which the
uncertainties reach an unacceptable level.

Although the effects of the higher dimension condensates are readily observed
in the figures, the inequalities are relatively insensitive to
the dimension-four gluon condensate and are virtually independent of the
possible (non-OPE) dimension-two contributions represented by $C_2$ this
implies that dimension-two phenomenological terms can be accommodated in
the sum-rules without violating the fundamental constraints imposed by
the H\"older inequalities.

Using H\"older's integral inequality we have constructed fundamental
constraints on the QCD sum-rules that must be satisfied if the sum-rule
is consistent with its phenomenological relation to the integral of
$Im\,\Pi(t)$.  As an example of the application of this idea, the
$s_0$, $\tau$ parameter space satisfying the inequalities was
determined for various choices of the condensates appearing in the
literature.
Except for the original analysis of the vector channel \cite{svz,svz2},
the parameters employed in the sum-rule analyses are inconsistent with the
inequalities.
Including a model for the uncertainties associated with truncation of the OPE
does not seem sufficient to account for this inconsistency.

We view the application of the inequalities as a practical method for
determining the energy ($\tau$) range over which the sum-rules are
valid.  In contrast to conventional approaches which rely upon estimates
of the uncertainties inherent in the sum-rules, the inequalities provide
a simple and fundamental constraint for studying the reliable energy
range of the QCD sum-rules.
Although one could devise models where the intrinsic errors associated with the
sum-rules are sufficiently large to accommodate violations of the inequalities,
we feel that the most
conservative approach  is to restrict a sum-rule analysis to regions of
parameter space where the inequalities are satisfied, rather that
relying upon error estimates (perhaps based on prejudice) to enforce
consistency of the sum-rules with the H\"older inequalities.
Furthermore, rough lower bounds on the
continuum threshold can be obtained, a result which is valuable in cases
where phenomenological estimates of the continuum are not available.

In conclusion, the H\"older inequalities for the
QCD sum-rules provides a useful and fundamental diagnostic for any sum-rule
analysis, and we encourage the use of this technique as a valuable consistency
check in any sum-rule application.

\bigskip\noindent
{\bf Acknowledgements:}
MB and TGS are grateful for the financial support of the Natural Sciences
and Engineering Research Council of Canada (NSERC).
GO thanks V. Vento for useful discussions.

\newpage

\begin{figure}
\caption{
The shaded area represents the  region
in the $s_0-\tau$ parameter space for the vector channel
consistent with the inequalities.
The values of the condensates indicated in the figure are taken
from Shifman-Vainshtein-Zakharov \protect\cite{svz,svz2} analysis
for the lower two graphs and twice vacuum saturation form  \protect\cite{nar}
for the upper graph.  The boxes represent the $s_0$-$\tau$ parameter space
used in the corresponding sum-rule analysis.
 The dashed line represents the
border of the parameter space when a dimension-two phenomenological
contribution
\protect\cite{domc2} is included.  The dotted line represents the border of
the parameter
space after modelling the effect of uncertainties in the power-law corrections.
}
\label{figure1}
\end{figure}

\begin{figure}
\caption{
Same as in Fig. \protect\ref{figure1} except for the axial-vector channel.
The lower two graphs correspond
to the parameters of \protect\cite{rry} and the upper graph represents the
parameters from \protect\cite{nar}.}
\label{figure2}
\end{figure}

\begin{figure}
\caption{The shaded area represents the region consistent with the inequalities
for both vector channel and axial-vector
channels using Dominguez-Sol\`a\protect\cite{dom} values for the condensates
up to and including dimension eight.
 The square boxes represent
the $s_0-\tau$ parameter space used in their analysis.
Dotted lines represent the effect of uncertainties as in
Fig. \protect\ref{figure1}.}
\label{figure3}
\end{figure}

\begin{figure}
\caption{The shaded region is consistent with the inequalities for both
vector channel and axial-vector channels
using Gim\`enez {\it et al.} \protect\cite{gim2} values for the condensates
up to dimension sixteen.
The square boxes represent the
$s_0-\tau$ parameter space as used in the  Gim\`enez
{\it et al.} \protect\cite{gim2} analysis.
As discussed in the text, dotted lines represent the effect of truncation of
the OPE at dimension 8.}
\label{figure4}
\end{figure}

\end{document}